\documentclass[preprintnumbers, prd, floatfix,twocolumn,
preprintnumbers, letterpaper, superscriptaddress,nofootinbib]{revtex4-1}
\usepackage{amsfonts}
\usepackage{amsmath}
\usepackage{amssymb,epsf}
\usepackage{latexsym}
\usepackage{graphicx,epsfig}
\usepackage{amssymb}
\usepackage{float}
\usepackage{subfigure}
\usepackage{epstopdf}
\usepackage[colorlinks=true,citecolor=blue,linkcolor=blue,urlcolor=black]{hyperref}
\usepackage{dcolumn}
\usepackage{psfrag}
\usepackage{wrapfig}
\usepackage{makeidx}
\usepackage{epsf}
\usepackage{color}
\usepackage{multirow}

\graphicspath{{Images/}}

\usepackage{tikz}

\definecolor{lime}{HTML}{A6CE39}
\newcommand{\orcidicon}{%
    \begin{tikzpicture}
    \draw[lime, fill=lime] (0,0)
        circle [radius=0.16]
        node[white] {{\fontfamily{qag}\selectfont \tiny ID}};
    \draw[white, fill=white] (-0.0625,0.095)
        circle [radius=0.007];
    \end{tikzpicture}   \hspace{-2mm}
}
\newcommand\orcidFrancisco{{\href{https://orcid.org/0000-0002-9388-8373}{\orcidicon}}}
\newcommand\orcidMahdi{{\href{https://orcid.org/0000-0003-1196-9493}{\orcidicon}}}

\begin{document}
\title{Observational optical constraints of regular black holes}

\author{Khadije Jafarzade}
\email{khadije.jafarzade@gmail.com}
\affiliation{ Department of Theoretical Physics, Faculty of Basic Sciences, University of Mazandaran, P. O. Box 47416-95447, Babolsar, Iran}
 \affiliation{ ICRANet-Mazandaran, University of Mazandaran, P. O. Box 47416-95447, Babolsar, Iran}
\author{Mahdi Kord Zangeneh\orcidMahdi\!\!}
\email{mkzangeneh@scu.ac.ir}
\affiliation{Physics Department, Faculty of Science, Shahid Chamran University of Ahvaz,
Ahvaz 61357-43135, Iran}
\author{Francisco S. N. Lobo\orcidFrancisco\!\!}
\email{fslobo@fc.ul.pt}
\affiliation{Instituto de Astrof\'isica e Ci\^encias do Espa\c{c}o, Faculdade de
Ci\^encias da Universidade de Lisboa, Edif\'icio C8, Campo Grande,
P-1749-016, Lisbon, Portugal}
\date{\today}

\begin{abstract}
In this work, we consider two recently introduced novel regular black hole solutions and investigate the circular null geodesics  to find the connection between the photon sphere, the horizon and the black hole shadow radii. We also study the energy emission rate for these solutions and discuss how  the parameters of models affect the emission of particles around the black holes. Furthermore,  we compare the resulting shadow of  these regular black holes with observational data of the Event Horizon Telescope and find the allowed regions of the model parameters for which the obtained shadow is consistent with the data. Finally, we employ the correspondence between the quasinormal modes in the eikonal limit and shadow radius to study the scalar field perturbations in these backgrounds.
\end{abstract}

\maketitle

\section{Introduction}

One of the most interesting predictions of general relativity (GR) is the existence of black holes,  which is confirmed by observational data \cite{LIGO2b}. Despite the fact that GR describes the spacetime outside the black hole event horizon consistently with observations within the statistical and systematic uncertainties \cite{LIGO2ba,Hees1a,LIGO3a,LIGO5a}, the theory contains shortcomings. For instance, in the strong field limit (at the Planck scale), some classical solutions in GR suffer from physical singularities which means that the theory breaks down in these regions. It should be noted that at high densities of matter, quantum effects become important and pressure may be able to counterbalance gravitational collapse. Thus, one could expect that when matter reaches the Planck density, the effects of quantum gravity kicks in and there may be enough pressure to prevent the formation of a singularity \cite{Lopezab}. This situation has been a motivation to study non-singular or regular black holes (RBHs). Historically, one of the first RBH solutions was found by Bardeen \cite{Bardeen1a}, who proposed a non-singular model of black holes by coupling GR minimally with nonlinear electrodynamics.
Subsequently, many black hole models with regular cores have been introduced \cite{Kumar1a,Dymnikova1ab,Bronnikov2ab,Balart2ab,Balart3ab,Neves4ab,Lobo:2020ffi}. RBHs have also been studied from various aspects including entropy \cite{Myung4m} and thermodynamics \cite{Myung4x,Aros2z,Kumar1y}, the quasilocal energy \cite{Balart2n}, the energy conditions \cite{Zaslavskiiac} and the inner horizon instability and unstable cores \cite{Rubio1x}.
Rotating RBHs \cite{Ghosh2zy,Held2x} and RBHs in higher-order curvature gravities \cite{Ghosh1ab,CANTATA:2021mgk} have also been investigated.

It seems that black holes are invisible, however, it is indeed possible to obtain an image of them. Recently, the Event Horizon Telescope (EHT) collaboration published an image of a supermassive black hole in M87 \cite{Akiyama1a,Akiyama1b,Akiyama1c,Akiyama:2019eap}. According to the observed image, there is a dark region surrounded by a bright ring which are called the black hole shadow and photon ring, respectively.
In fact, the shadow and photon ring arise as a result of gravitational lensing in a strong gravity regime. The shadow of a black hole is an observational tool to understand the fundamentals of gravitation theory and provides relevant information concerning jets and matter dynamics around these compact objects. The black hole shadow can also be interpreted as a source of data describing the black hole parameters, such as the mass, charge and rotation. However, massless photons are the most convenient test particles for observational purposes, as they can form a photon sphere in the exterior of the black hole horizon. 

The photon sphere is typically unstable and can cause a ``shadow'' for faraway observers. Although the intensity map of an image depends on the details of the emission mechanisms of photons, the contour of the shadow depends only on the spacetime metric itself, since it corresponds to the apparent shape of the unstable light rings. Thus, strong lensing images and shadows not only help us to detect the nature of a compact object but also provide an opportunity to test whether the gravitational field around a compact object is described by the Schwarzschild or Kerr geometry. Motivated by specific interesting properties of the photon spheres and shadows, much research has been undertaken, including the seminal works of Refs. \cite{Synge1a,Luminet1a}, the shadow of a Schwarzschild black hole and extensions to rotating cases \cite{Vries1a,Kenta1a,Atamurotov1c,Atamurotov1a,Atamurotov1d,Perlick1a,Cunha1a}, optical properties of a braneworld black hole \cite{Atamurotov1h}, black holes in modified gravity \cite{Atamurotov1b,Atamurotov1e,Atamurotov1g}, and gravitational lensing in the presence of plasma \cite{Atamurotov1f,Atamurotov2lp,Atamurotov1ii,Atamurotov1jj,Atamurotov2k}, amongst others. In particular, different aspects of the Kerr black hole shadow such as the measurement of the mass and spin parameter have been analysed in Refs. \cite{Kerrshadow1a,Kerrshadow1b,Kerrshadow1c,Kerrshadow1d,
Kerrshadow1e,Kerrshadow1f,Kerrshadow1g,Kerrshadow1h}. Recently, the light rings and shadows of an uniparametric family of spherically symmetric geometries interpolating between the Schwarzschild solution, a regular black hole, and a traversable wormhole, and dubbed as black bounces, all of them sharing the same critical impact parameter, were extensively studied in Ref. \cite{Guerrero:2021ues}. More specifically, the ray tracing method was considered in order to study the impact parameter regions corresponding to the direct, lensed, and photon ring emission, and it was found that there is a broadening of all these regions for the black bounce solutions as compared to the Schwarzschild one.

An interesting novel model of RBHs which has recently attracted much attention is the Simpson-Visser black-bounce geometry \cite{Nascimento1a,Tsukamoto1a,Bronnikov1a,Tsukamoto1b}. In Refs. \cite{Haroldo1bd,Lima1a}, it was shown that the Simpson-Visser black-bounce is shadow-degenerate with Schwarzschild. In this paper, we first consider a charged black bounce  and check whether such a condition can be satisfied for charged cases in  the Simpson-Visser spacetime as well. In the following, we take into account another RBH with asymptotically Minkowski core \cite{Simpson:2019mud,Berry} and show that although in the Simpson-Visser black-bounce geometry, the shadow radius is independent of its model parameter, one can find other RBHs in which the shadow radius is governed by the parameters of the model. We then compare the results to the shadow of the supermassive black hole at the center of M87 imaged by the EHT collaboration. This enables us to make some novel constraints on the parameters of the models. Furthermore, we also study the energy emission rate and quasinormal modes (QNMs) for these RBHs.

The plan of the paper is as follows:  In Sec. \ref{I}, we introduce the charged black-bounce RBH model, review its properties and discuss the corresponding null geodesics. Then, we study the optical features including the geometrical shape of the shadow and the energy emission rate for this RBH in Subsec. \ref{sh&ph} and discuss how the model parameters affect these quantities. In Subsec. \ref{III}, we compare the resulting shadows to the one detected by the EHT collaboration and study the constraints on the parameters of the model. In Subsec. \ref{QNMI}, we employ the correspondence between QNMs and the black hole shadow in the eikonal limit to find the quasinormal frequencies under test scalar field perturbations. In Sec. \ref{RBII}, we consider another RBH with asymptotically Minkowski core and investigate the mentioned discussions for this solution.  Finally, we discuss and summarize our results in Sec. \ref{V}.

\section{Charged black-bounce and null geodesics}\label{I}

We first consider an interesting novel model of a RBH, recently proposed \cite{Franzin1ad}, for which the general form of the metric is
\begin{equation}
ds^{2}=-A(r)dt^{2}+B(r)dr^{2}+C(r)(d\theta^{2}+\sin^2\theta \; d\phi^2),
\label{EqBbounce11}
\end{equation} 
 where $A(r)$, $B(r)$, and $C(r)$ are given by
\begin{eqnarray}\label{EqAr}
A(r) = \frac{1}{B(r)}\equiv 1-\frac{2m}{\sqrt{r^{2}+a^{2}}}+\frac{Q^{2}}{r^{2}+a^{2}},~C(r) = r^{2}+a^{2}, \notag \\
\end{eqnarray}
in which $m$ is the total mass, $ Q $ and $a$ are the electric charge and bounce parameter, respectively. Note that the Reissner–Nordstrom solution is obtained for $a=0$. This structure models a RBH with a horizon located  at 
\begin{equation}
r_{e} = \sqrt{2m^{2}-Q^{2}-a^{2}+2m\sqrt{m^{2}-Q^{2}}}.
\label{Eqre}
\end{equation} 
For $ a >\sqrt{2m^{2}-Q^{2}+2m\sqrt{m^{2}-Q^{2}}}$, no physical solution can be observed for the black bounce.

In order to investigate the optical properties of this black hole, we first find how light rays move around the black hole. To this purpose, we employ the geodesic equation to calculate the radius of the innermost circular orbit for a test particle in the background of the black hole. The geodesic motion of a massless photon, in the static spherically symmetric spacetime, can be obtained by the following Hamiltonian
\begin{equation}
H=\frac{1}{2}g^{ij}p_{i}p_{j}=0.
\label{EqHamiltonian}
\end{equation}
Due to the spherically symmetric characteristics of these black holes, without a loss of generality we consider trajectories of photons on the equatorial plane with $\theta=\pi/2$.
Thus, Eq. (\ref{EqHamiltonian}) can be written as
\begin{equation}
\frac{1}{2}\left[-\frac{p_{t}^{2}}{A(r)}+\frac{p_{r}^{2}}{B(r)}+\frac{p_{\phi}^{2}}{C(r)}\right] =0,
\label{EqNHa}
\end{equation}
from which we deduce
 \begin{equation}
\dot{p_{t}}= -\frac{\partial H}{\partial t}=0, \qquad  \dot{p_{\phi}}= -\frac{\partial H}{\partial \phi}=0,
\label{Eqpt1}
\end{equation}
so that one can consider $ p_{t} $ and $ p_{\phi} $ as  constants of motion. We define $ p_{t}=-E$ and $ p_{\phi}=L $, where  $ E $ and $ L $ are the energy and angular momentum of the
photon, respectively. 

The equations of motion for photons are obtained as
\begin{eqnarray}
\dot{t}&=&\frac{\partial H}{\partial p_{t}}=-\frac{p_{t}}{A(r)}, \;\;
\dot{r}=\frac{\partial H}{\partial p_{r}}=\frac{p_{r}}{B(r)}, \;\;
\dot{\phi}=\frac{\partial H}{\partial p_{\phi}}=\frac{p_{\phi}}{C(r)},
\nonumber
\end{eqnarray}
where the overdot is the derivative with respect to the affine parameter and $ p_{r} $ is the radial momentum. These equations provide a complete description of the radial motion as
\begin{equation}
\dot{r}^{2}+V_{\rm eff}(r)=0, 
\label{EqVef1}
\end{equation}
where $ V_{\rm eff} $ is the effective potential of the photon, given by
\begin{equation}
V_{\rm eff}(r)= \frac{1}{B(r)}\left[ \frac{L^{2}}{C(r)}-\frac{E^{2}}{A(r)}\right].
\label{Eqpotential}
\end{equation}

\begin{figure}[t!]
\includegraphics[width=0.3\textwidth] {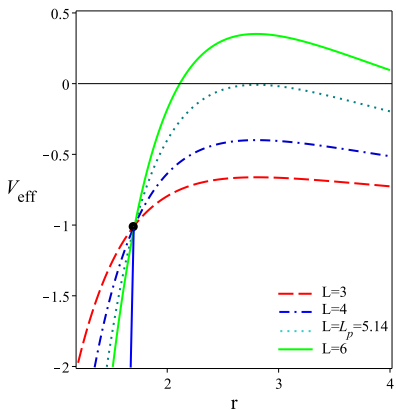}
\caption{The behavior of effective potential $ V_{\rm eff}(r) $ for $ E=m=1 $, $Q=0.2$ and bounce parameter $a=1$ with different values of $ L $. The blue solid line is related to the place of the horizon where $ V_{\rm eff}=-1 $.}
\label{FigVef}
\end{figure}
The behavior of the photon's effective potential for different values of $L$ is illustrated in Fig. \ref{FigVef}. As we see the maximum of the potential increases as $L$ increases. Since $ \dot{r}^{2} \geq 0 $, we expect that the effective potential satisfies $ V_{\rm eff}\leq 0 $. Thus, the photon trajectories can only appear for a negative effective potential. For lower values of $L$, an ingoing photon can fall into the black hole, whereas for larger values of $L$, it will be reflected before it falls into the black hole. An interesting phenomenon is related to the critical angular momentum $L = L_{p}$ for which $ \max (V_{\rm eff}) = 0$. In this case, the photon has zero radial velocity at $ \max (V_{\rm eff})$ and orbits the black hole due to its non-vanishing transverse velocity. For a spherically symmetric static black hole, it corresponds to the photon sphere. From what was expressed, one can find that the photon orbits are circular and unstable associated to the maximum value of the effective potential. Such a maximum value can be obtained by the following conditions 
\begin{equation}
V_{\rm eff}(r_{ph})=0,\qquad 
V^{\prime}_{\rm eff}(r_{ph})=0, \qquad
V^{\prime \prime}_{\rm eff}(r_{ph})<0 ,\label{EqVeff1}
\end{equation}
where the first two conditions determine $ L_{p} $ and the photon sphere radius ($ r_{ph} $) respectively, while the third condition ensures that the photon orbit is unstable. Hereafter, we will fix $E$ and $m$ to unity, for simplicity, in our calculations.

\subsection{Photon sphere, shadow and energy emission rate}\label{sh&ph}

In this subsection, we discuss the photon sphere, shadow and energy emission rate of the charged black bounce RBH. In the geometric optics limit, one can consider the motion of a photon as a null geodesic. We calculate the radius of the photon sphere and shadow for the black bounce solution by solving Eqs. (\ref{EqVeff1}). Furthermore, we study the energy emission rate of each solution.

Taking into account  the effective potential (\ref{Eqpotential}) and conditions (\ref{EqVeff1}), $V^{\prime}_{\rm eff}(r_{ph})=0$ gives $ C^{\prime}(r_{ph}) A(r_{ph})-A^{\prime}(r_{ph}) C(r_{ph})=0 $ and using the metric functions (\ref{EqAr}), we have 
\begin{equation}
X^{2}-3mX+2Q^{2}=0,
 \label{Eqrph2}
\end{equation}
where $ X=\sqrt{r_{ph}^{2}+a^{2}}$. Solving Eq. (\ref{Eqrph2}), one can obtain the following relation
\begin{equation}
X=\frac{1}{2}\left(3m+\sqrt{9m^{2}-8Q^{2}}\right)=r_{ph}^{(\mathrm{RN})},
 \label{Eqrph3}
\end{equation}
which shows that $X$ is equal to photon radius of the Reissner-Nordstrom black hole, $r_{ph}^{(\mathrm{RN})}$. It is evident that the condition $ Q<\frac{3}{2\sqrt{2}} m$ should be satisfied in order to have real $X$.

The corresponding constant of motion $L/E$ for this photon sphere is obtained by $V_{\rm eff}(r_{ph})=0$ and is given as \cite{Masmar}
\begin{equation} 
\label{Eqshadow1}
r_{sh}=\frac{L_{p}}{E}=\sqrt{\frac{C(r_{ph})}{A(r_{ph})}},
\end{equation}
where $r_{sh}$ is the radius of shadow. Thus, using Eq. (\ref{Eqshadow1}), the shadow radius takes the form
\begin{equation} 
r_{sh}=\frac{X^{2}}{\sqrt{Q^{2}-2mX+X^{2}}}.
\label{EqrshII}
\end{equation}
The above equation shows automatically that the shadow radius is the same for the Reissner-Nordstrom and Simpson-Visser geometries.
As we see, although the radius of the photon sphere depends on the bounce parameter, the shadow radius is independent of this parameter. This reconfirms the results of \cite{Lima1a,Haroldo1bd} for uncharged black bounce geometry and shows that even in the presence of electric charge, the shadow radius of Simpson-Visser black-bounce is independent of its model parameter.
\begin{figure}[t!]
\centering
 \includegraphics[width=0.39\textwidth]{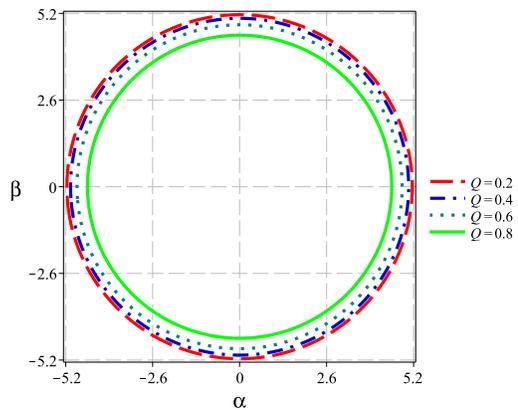} 
\caption{The black bounce shadow in the celestial plane  $ (\alpha - \beta) $ for $ m = 1 $.}
\label{Fig7}
\end{figure}
\begin{figure*}[t!]
\subfigure[~$ Q=0.2 $]{
   \label{FigEr2}  \includegraphics[width=0.31\textwidth] {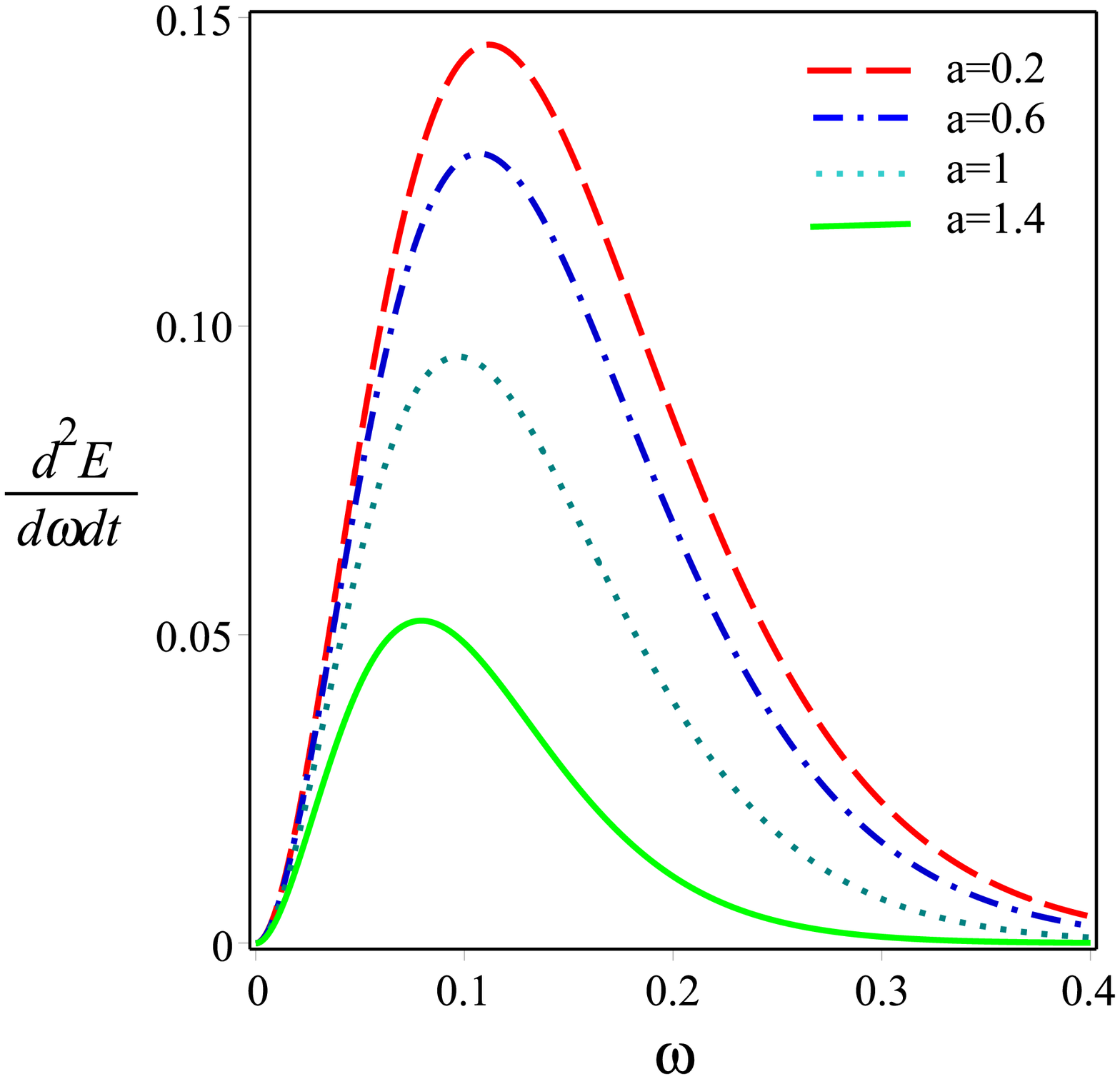} }\qquad \qquad \qquad \qquad 
   \subfigure[~$a=1 $]{
   \label{FigEr3}  \includegraphics[width=0.31\textwidth] {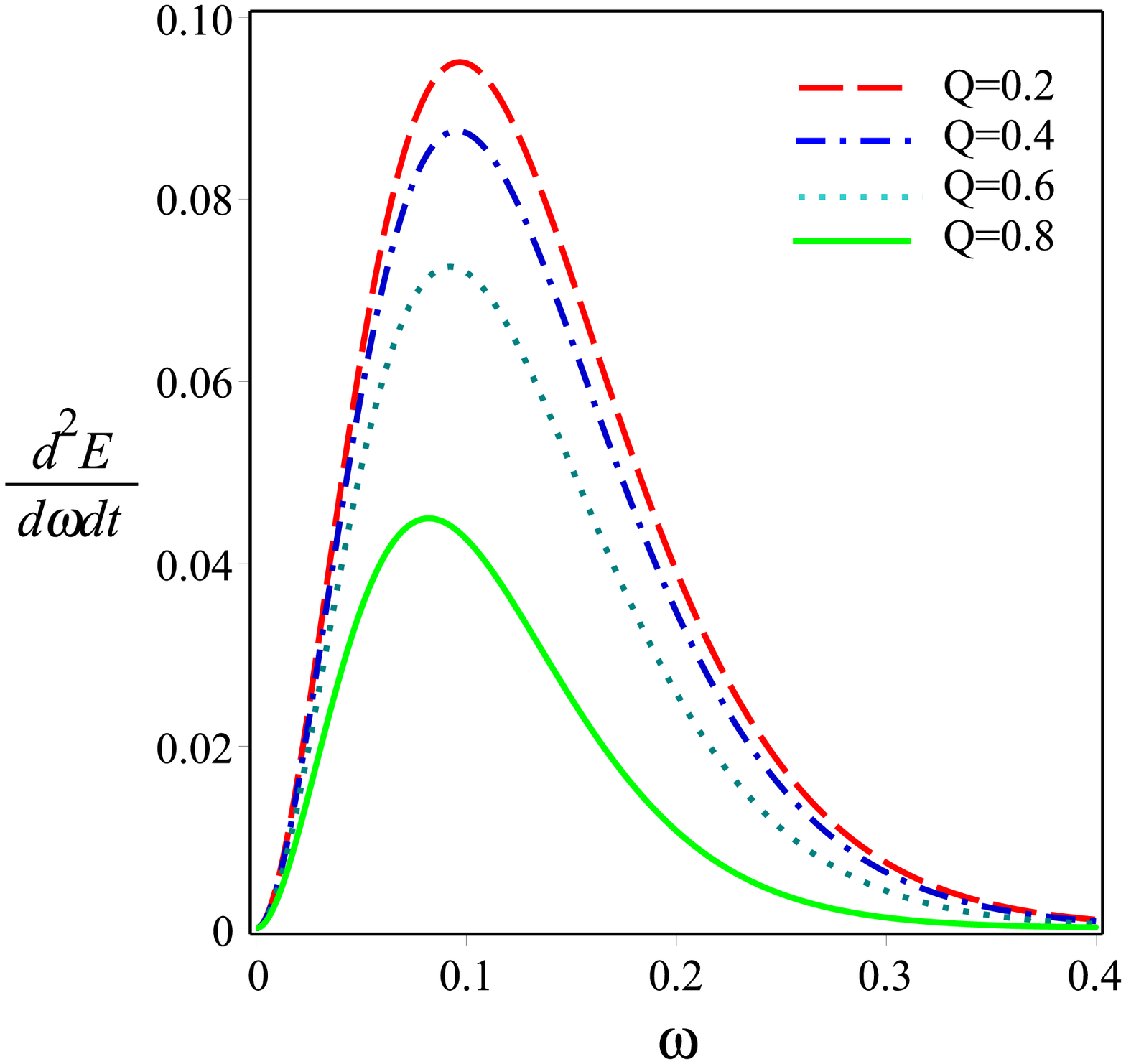} }
\caption{The energy emission rate versus $\omega$ for the charged black bounce with $m = 1$ and different values of bounce parameter and electric charge.}
\label{FigEr}
\end{figure*}

It is worthwhile to mention that in Ref. \cite{HLu}, the authors found the following inequalities between several parameters characterizing the black hole size:
\begin{equation}
\frac{3}{2}r_{e}\leq r_{ph}\leq \frac{r_{sh}}{\sqrt{3}}\leq 3m.
\label{Eqr}
\end{equation}
Applying these inequalities (\ref{Eqr}) to the the Simpson-Visser black-bounce geometry, according to our investigations, these inequalities are preserved for all values of the bounce parameter $ a $. 

Here, we explore the influence of the electric charge and bounce parameter on the optical quantities. As we have already mentioned, the shadow radius depends only on the electric charge for a fixed mass parameter. This effect is depicted in Fig. \ref{Fig7} which displays the decreasing contribution of the electric charge on the shadow radius.

Next, we investigate the emission of particles around this black hole solution.
It was shown that for a far distant observer, the absorption cross-section corresponds to the black hole shadow \cite{WWei,Belhaj1,Belhaj2}. In general, at very high energies, the absorption cross-section for spherically symmetric black holes oscillates around a limiting constant value $ \sigma_{lim} $ which is approximately equal to the area of the photon sphere ($ \sigma_{lim} \approx \pi r_{sh}^{2}$). The energy emission rate can be obtained by the following relation
\begin{equation}
\frac{d^{2}\mathcal{E}(\omega)}{dtd\omega}=\frac{2\pi^{3}\omega^{3}r_{sh}^{2}}{e^{\frac{\omega}{T}} -1},
\label{Eqemission}
\end{equation}
where $ \omega $ is the emission frequency and $ T $ is the Hawking temperature.
For the charged black bounce, the Hawking temperature is given by 
\begin{equation}
T=\frac{r_{e}}{2\pi}\frac{m \sqrt{a^{2}+r_{e}^{2}}-Q^{2}}{(a^{2}+r_{e}^{2})^{2}}.
\label{EqTHI}
\end{equation}

Figure \ref{FigEr} depicts the behavior of energy emission rate as a function of $\omega$. Taking a close look at Figure \ref{FigEr2}, one can find that the maximum value of the emission rate decreases and shifts to low frequencies as the bounce parameter increases. In fact, reducing this parameter leads to a fast emission of particles. Regarding the effect of electric charge, as we see from Fig. \ref{FigEr3}, it has a decreasing effect on this quantity similar to the bounce parameter.

\subsection{Comparison with the EHT data} \label{III}
In this subsection, we compare the resulting shadows for the charged black-bounce RBH with the one detected by the Event Horizon Telescope (EHT) \cite{Akiyama1a}. As we showed, the shadow size depends on the electric charge and, consequently, this parameter should be constrained in order to have a result which is consistent with the EHT observation.

According to the obtained results in Ref. \cite{Akiyama1a}, the angular size of the shadow of M87* is $ \delta = (42 \pm 3) \mu as $. Following \cite{Akiyama:2019eap}, the distance to M87* is obtained as $ D = 16.8^{+0.8}_{-0.7} Mpc$, while the mass of the object is $M = (6.5 \pm 0.9) \times 10^{9} M_{\odot}$ with $M_{\odot}$ the mass of the Sun. These numbers imply that the diameter of the shadow in units of mass should be \cite{bambi,EHTM8}
\begin{equation}
d_{M87^{*}}\equiv \frac{D\delta}{M}\approx 11.0 \pm 1.5.
\label{EqdM87}
\end{equation}
As we see from Eq. (\ref{EqdM87}),  within $1\sigma $ uncertainty $ 9.5\lesssim d_{M87^*} \lesssim 12.5$  whereas within $2\sigma $ uncertainty $ 8.0 \lesssim d_{M87^*} \lesssim 14.0$.

\begin{figure}[t!]
 \includegraphics[width=0.31\textwidth] {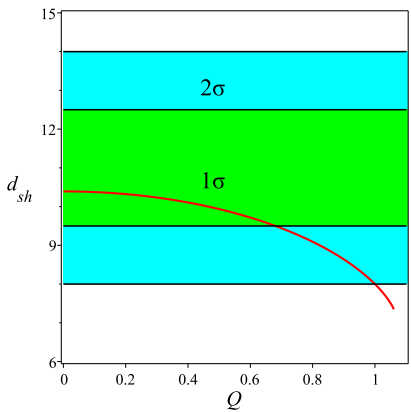}          
\caption{ Diameter of the shadow ($d_{sh}$) of the charged black bounce as a function of the
electric charge. The shaded regions indicate the
values of $d_{sh}$  consistent with the EHT data. The green shaded region gives the $ 1\sigma $ confidence region for $d_{sh}$, whereas the
cyan shaded region gives the $ 2\sigma $ confidence region.}
\label{Fig12}
\end{figure}

In Fig. \ref{Fig12} we have plotted the diameter of the resulting shadow as a function of the electric charge together with $ 1\sigma $ and $ 2\sigma $ confidence intervals on the diameter of the shadow of M87*. From this figure it is evident that one can set an upper limit on the electric charge such that if this quantity increases too much, the size of the shadow is inconsistent with observations. From this parameter scan, one can also find a rough upper limit of $Q \leq 0.68 $ at $ 1\sigma $ and $Q \leq 1.0 $ at $ 2\sigma $. 

\subsection{Quasinormal modes}\label{QNMI}

In this subsection, we employ the shadow results to investigate one of the interesting dynamic properties of black holes, namely, quasinormal modes (QNMs). In fact, black holes always interact with matter and fields around them and as a result of these interactions, they take a perturbed state.
When black holes are perturbed, they tend to relax towards equilibrium. Such a process is done through the emission of QNMs.

Here, we consider the propagation of a massless scalar field $  \Phi$ in a fixed gravitational background. The equation of motion for $\Phi$ is then given by the usual Klein-Gordon equation $\nabla^{\mu}\nabla_{\mu}\Phi=0$. In order to solve this equation, one can apply the method of separation of variables as follows
\begin{equation}
\Phi (t, r, \theta, \phi)=\frac{\psi(r)}{r}e^{-i\omega t}Y_{lm}(\theta, \phi),
\end{equation}
where $ Y_{lm}(\theta, \phi) $ are the standard spherical harmonics, and $ \omega $ is  the frequency of the perturbation. Using the above ansatz,  it is straightforward to obtain the radial part of the wave equation which is a second order differential equation
 \begin{equation}
\frac{d^{2}\psi}{dx^{2}}+ \left[ \omega^{2} - V(x) \right]\psi=0,
\end{equation}
where $x\equiv \int dr/{A(r)}$ is the tortoise coordinate. Finally, the effective potential for scalar perturbations is given by
\begin{equation}
V_{s}(r)=A(r)\left( \frac{l(l+1)}{r^{2}}+\frac{A^{\prime}(r)}{r}\right) ,
\label{EqEV2}
\end{equation}
where the prime denotes the derivative with respect to $r$, and $ l\geq 0 $ is the angular number.

Over the years, several different methods have been suggested to compute the QNMs of black holes such as the WKB approximation \cite{Schutz2m}, series solutions in asymptotically AdS backgrounds \cite{Horowitz3a}, Leaver’s continued fraction method \cite{Starinets3a} and the monodromy methods \cite{Motl1c}. Among different methods, semi-analytical methods based on WKB approximation are the most popular ones due to sufficient accuracy \cite{Fernando3b,Fernando4ab,Fernando2ac,Kokkotas1,Fernando1}. Within the WKB approximation, the QN frequencies are given by
\begin{equation}
\omega_{n}^{2}=V_{0}+\sqrt{-2V_{0}^{\prime \prime}}\Lambda (n) -i\nu \sqrt{-2V_{0}^{\prime \prime}}\left[ 1+\Pi(n)\right] 
\label{EqEV},
\end{equation}
where $ \nu = n + 1/2 $ with $ n = 0, 1, 2... $ is the overtone number. $ V_{0} $ represents the height of the barrier, namely, the maximum of the potential and $  V_{0}^{\prime \prime}$ is the second derivative of the potential evaluated at the maximum. $ \Lambda (n) $ and $ \Pi (n) $ are complicated expressions of $ n $ and higher derivatives of the potential evaluated at the maximum. Their precise forms can be found in Refs. \cite{Kokkotas1,Fernando1}.

Here we study the effects of the model's parameters on the scalar field perturbation by using the correspondence between QNMs in the eikonal limit and shadow radius \cite{Jusufi1a,Jusufi1b}. In the eikonal regime, the WKB approximation becomes far accurate, hence the QN spectrum can be computed within the first order WKB semi-analytical approach. In such a limit the angular momentum term would be the dominant one in the effective potential barrier. In this limit, 
QNM frequencies can be obtained by the following relation \cite{VCardoso1}
\begin{equation}
\omega = l \, \Omega -i \left(n+\frac{1}{2} \right) | \lambda |,
\label{EqQNM1}
\end{equation}
where $ \Omega $ is the coordinate angular velocity given as,
\begin{equation}
\Omega =\frac{\dot{\phi}}{\dot{t}}=\sqrt{\frac{A(r_{ph})}{C(r_{ph})}}=\frac{1}{r_{sh}},
\label{EqQNM2}
\end{equation}
and the Lyapunov exponent $ \lambda $ is interpreted as the decay rate of the unstable circular null geodesics and is expressed as
\begin{equation}
\lambda =\sqrt{\frac{-V_{\rm eff}^{\prime \prime}(r_{ph})}{2\dot{t}^{2}}}.
\label{EqQNM3}
\end{equation}
Note that for spherically symmetric black holes, $ r_ {ph} $ matches the place in which the potential is maximum. Taking into account Eqs. (\ref{EqQNM2}) and (\ref{EqQNM3}), we are in a position to probe how the model's parameters affect QNMs.
\begin{figure*}[t!]
\subfigure[]{
   \label{FigQ3}  \includegraphics[width=0.32\textwidth] {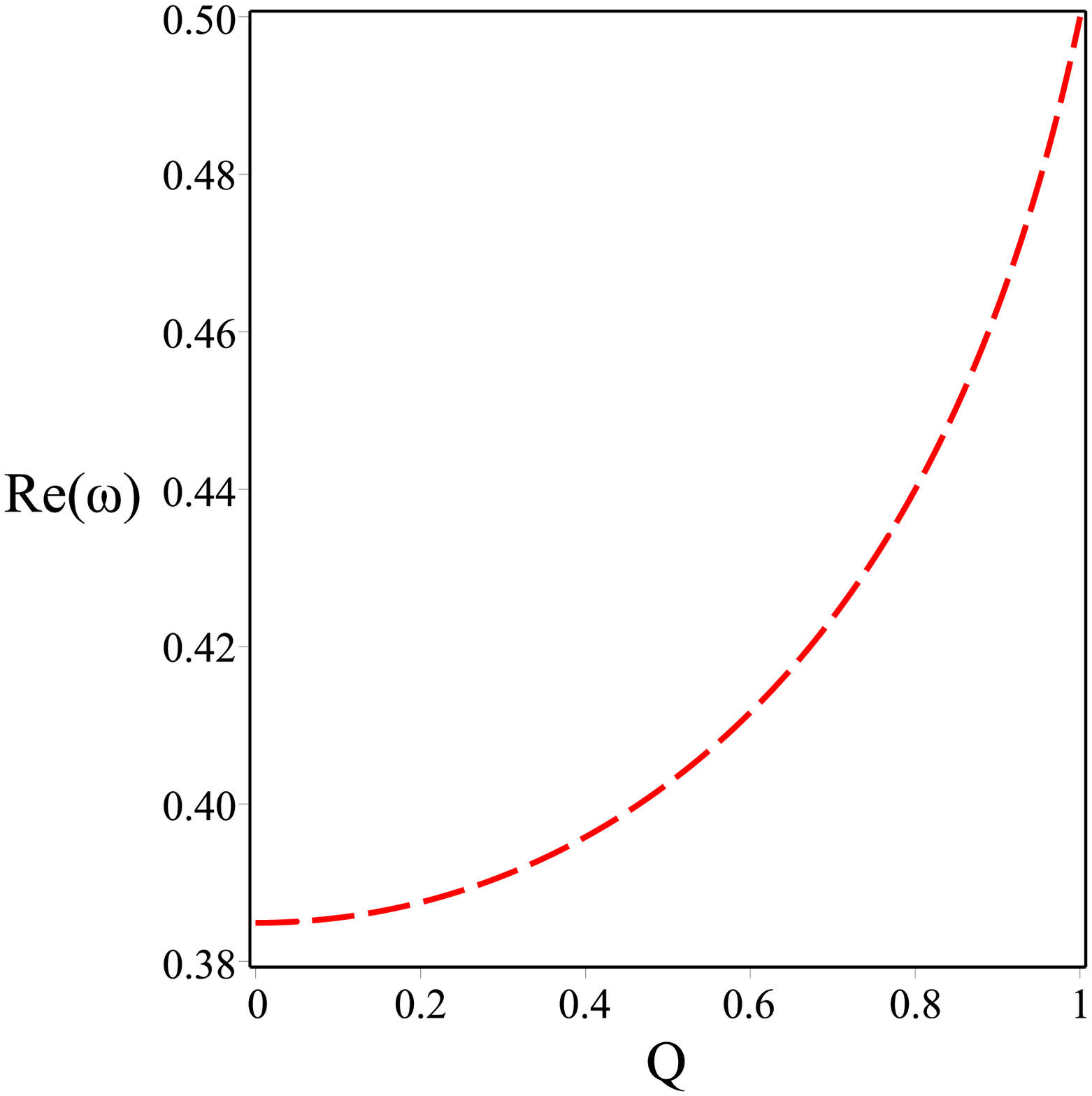} }\qquad \qquad \qquad \qquad 
 \subfigure[]{
   \label{FigQ4}  \includegraphics[width=0.32\textwidth]{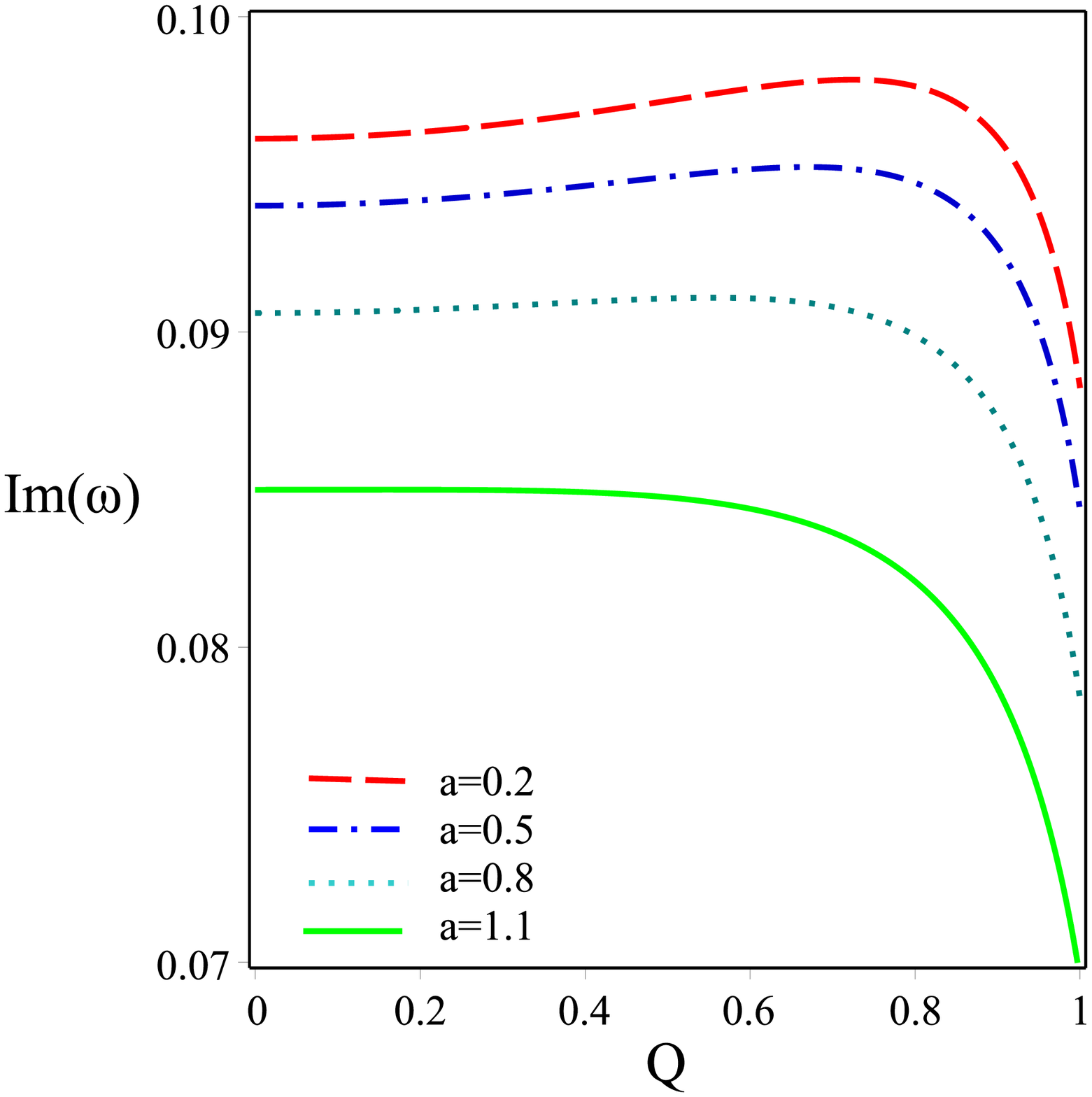}}
\caption{The behavior of Re($\omega$) and  Im($\omega$) with respect to electric charge for $ m = 1 $, $n = 0$, $l = 2$ and different values of the bounce parameter.}
\label{Fig10}
\end{figure*}
Now, we would like to study the QN frequencies around the black bounce. According to Eq. (\ref{EqQNM2}), the real part of the QNM frequency Re($\omega$) depends only on the electric charge, whereas the imaginary part Im($\omega$) is dependent on both parameters (see Eq. (\ref{EqQNM3})). Fig. \ref{FigQ3} gives a simple illustration of the influence of electric charge on Re($\omega$). Clearly, Re($\omega$) is an increasing function of the electric charge. This indicates that the scalar field perturbations in the presence of electric charge oscillate faster as compared to neutral black holes. Regarding the effect of this parameter on Im($\omega$), taking a close look at Fig. \ref{FigQ4},  one can find that depending on the value of the bounce parameter $a$, increasing $Q$ leads to the increasing or decreasing the imaginary value of the QNM frequency. For larger values of $a$, increasing $Q$ makes the decreasing of Im($\omega$), indicating that the scalar field perturbations decay slower under such conditions. For smaller values of $a$, all curves increase firstly with the growth of $Q$, then they gradually decrease by increasing this parameter. In other words, they have a global maximum value. This means that there exists a finite value of the electric charge where the scalar field perturbations decay very fast.  Fig. \ref{FigQ4} also shows the influence of the bounce parameter on Im($\omega$), which reveals the fact that the existence of a smaller bounce parameter makes the perturbations decay faster.

\section{RBH with asymptotically Minkowski core}\label{RBII}
 
Although for the Simpson-Visser black-bounce geometry the shadow radius is independent of its model parameter, one can find other regular black holes for which the shadow radius is governed by the model's parameters. In this section, we are interested in considering another RBH with an exponentially suppressing mass (in the asymptotically Minkowski core region) controlled by a suppression parameter.

The metric functions for this black hole are \cite{Simpson:2019mud,Berry}
\begin{equation}
A(r) = \frac{1}{B(r)}\equiv 1-\frac{2m e^{-\frac{a}{r}}}{r},~~C(r) = r^{2},
\label{metfunct1}
\end{equation}
where $ m $ and $ a $ are, respectively, the mass and the suppression parameter. Evidently, for $ a=0 $, this solution reduces  to the Schwarzschild case. For $0 < a \leq 2m/e$, we have physical black hole solutions for which horizons are given by \cite{Simpson:2019mud,Berry}
\begin{equation}
r_{e}=2me^{W(-\frac{a}{2m})} ,
\label{EqrootMin}
\end{equation}
where $W(x)$ is the real-valued Lambert $ W $ function.

\begin{figure}[t!]
 \includegraphics[width=0.32\textwidth] {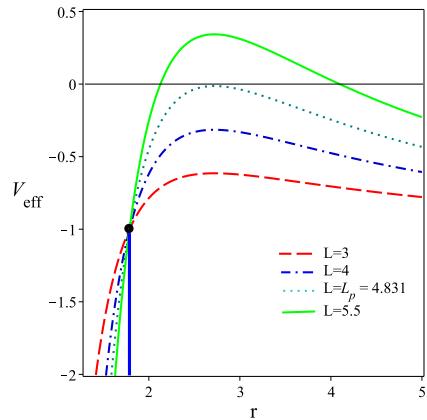}
\caption{The behavior of effective potential $ V_{\rm eff}(r) $ for $ E=m=1 $ and suppression $a=0.2$ with different values of $ L $.}
\label{FigVefII} 
\end{figure}

Now, we calculate the optical quantities mentioned in the previous section for this model as well, and explore the effect of suppression parameter on them. To do so, we first calculate $V_{\rm eff}$ through Eq. (\ref{Eqpotential}) and investigate its behavior which is illustrated in Fig. \ref{FigVefII}. According to this figure, the effective potential has positive values for $L> 4.831$ which is physically unacceptable.

In order to calculate the radius of photon sphere, we employ  $V^{\prime}_{\rm eff}(r_{ph})=0$ which leads to the following  relation
\begin{equation}
 2r_{ph}^{2}e^{\frac{a}{r_{ph}}}-6mr_{ph}+2ma=0.
 \label{Eqrph}
\end{equation}

Using Eq. (\ref{Eqshadow1}), we can calculate the
shadow radius as
\begin{equation} 
r_{sh}^{2}=\frac{L_{p}^{2}}{E^{2}}=\frac{r_{ph}^{2}}{A(r_{ph})}.
\label{Eqrsh}
\end{equation}

As we see, Eq. (\ref{Eqrph}) is not solvable exactly and has to be solved numerically. In table \ref{tab1},  we list the numerical solutions of Eqs. (\ref{Eqrph}) and (\ref{Eqrsh}) as well as the event horizon radius $r_e$, for various suppression parameters and compare our results to those of the Schwarzschild solution given by $a=0$. We verify that $ r_{e} $, $ r_{ph} $ and $ r_{sh} $ are decreasing functions of $ a $. To investigate the inequality relations
(\ref{Eqr}),  from data of table \ref{tab1}, one could check that these relations are preserved for the corresponding black hole solution. 

\begin{table}[h]
\begin{center}
\begin{tabular}{|c| c| c| c| c| c| c| c| c|c|c|}
\hline\hline
$~a~$&$~r_{e}~$&$~r_{ph}~$&$~r_{sh}~$&$~\frac{r_{ph}}{3r_{e}/2}~$&$~\frac{r_{ph}}{3m}~$&$~\frac{r_{sh}}{\sqrt{3}r_{ph}}~$&$~\frac{r_{sh}}{3\sqrt{3}m}~$ \\ \hline\hline
$~0.0~$&$~2.0~$&$~3.0~$&$~5.19~$&$~1.0~$&$~1.0~$&$~1.0~$&$~1.0~$ \\ \hline
$~0.1~$&$~1.89~$&$~2.86~$&$~5.01~$&$~1.008~$&$~0.95~$&$~1.012~$&$~0.96~$ \\ \hline
$~0.2~$&$~1.78~$&$~2.71~$&$~4.83~$&$~1.014~$&$~0.90~$&$~1.029~$&$~0.92~$ \\ \hline
$~0.24~$&$~1.74~$&$~2.65~$&$~4.75~$&$~1.015~$&$~0.88~$&$~1.030~$&$~0.91~$ \\ \hline
$~0.3~$&$~1.67~$&$~2.56~$&$~4.63~$&$~1.021~$&$~0.85~$&$~1.044~$&$~0.89~$ \\ \hline
$~0.4~$&$~1.54~$&$~2.39~$&$~4.21~$&$~1.034~$&$~0.79~$&$~1.071~$&$~0.81~$ \\ \hline
$~0.5~$&$~1.39~$&$~2.21~$&$~4.18~$&$~1.059~$&$~0.73~$&$~1.092~$&$~0.79~$ \\ \hline
$~0.57~$&$~1.28~$&$~2.06~$&$~4.00~$&$~1.075~$&$~0.68~$&$~1.121~$&$~0.77~$ \\ \hline
$~0.6~$&$~1.22~$&$~2.03~$&$~3.92~$&$~1.094~$&$~0.66~$&$~1.116~$&$~0.75~$ \\ \hline
$~0.7~$&$~0.97$&$~1.73~$&$~3.61~$&$~1.183~$&$~0.57~$&$~1.203~$&$~0.69~$ \\ \hline
\hline
\end{tabular}
\end{center}
\caption{The values of event horizon $r_e$, photon sphere $r_{ph}$ and shadow $r_{sh}$ radii and some combinations of them relating to the inequality relations (\ref{Eqr}), for different values of suppression parameter $a$ with $m=1$.}
\label{tab1}
\end{table}

As we discussed in the previous section, the energy emission rate could be obtained by Eq. (\ref{Eqemission}) in which $T$ is the Hawking temperature given by
\begin{equation}
T=\frac{me^{-\frac{a}{r_{e}}}(1-\frac{a}{r_e})}{\pi r_{e}^{2}}.
\label{EqTHI2}
\end{equation}

\begin{figure}[h!]
\centering
   \includegraphics[width=0.35\textwidth] {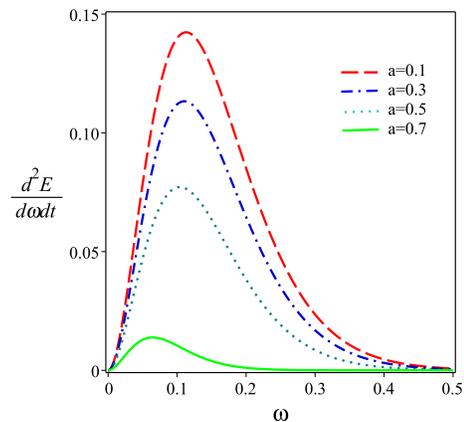}
\caption{The energy emission rate versus $\omega$ for RBHs with asymptotically Minkowski core with $m=1$ and different values of suppression parameter.}
  \label{FigEr1}
\end{figure}

In Fig. \ref{FigEr1}, the behavior of the energy emission rate with respect to $ \omega $ is displayed. As one can see, there exists a peak for the energy emission rate which decreases by increasing the suppression parameter $ a $ and shifts to the low frequencies with growth of this parameter. In other words, a decrease of $a$ implies a fast emission of particles. This shows that the black hole has a shorter lifetime in the presence of a smaller suppression parameter.

\subsection{Comparison with the EHT data}

Here we examine the compatibility of the resulting shadow of RBHs with asymptotically Minkowski core with EHT data.
As it was mentioned, the shadow size for such black holes is a decreasing function of $ a $. Since increasing this parameter leads to a smaller diameter of the shadow (see table \ref{tab1}), we can expect to set an upper limit on $ a $ to have a consistent result with the EHT detection. Taking a close look at table \ref{tab1}, one can find that  $ 9.5 <d_{sh}=2r_{sh} < 12.5$ ($1\sigma $ confidence interval) occurs in the range of $ 0 <a< 0.24$. Within the $2\sigma$ uncertainty, namely, $ 8.0 <d_{sh}< 14.0$, the acceptable range is $ 0 <a< 0.57$.
\begin{figure*}[t!]
\subfigure[]{
   \label{FigQ1}  \includegraphics[width=0.33\textwidth] {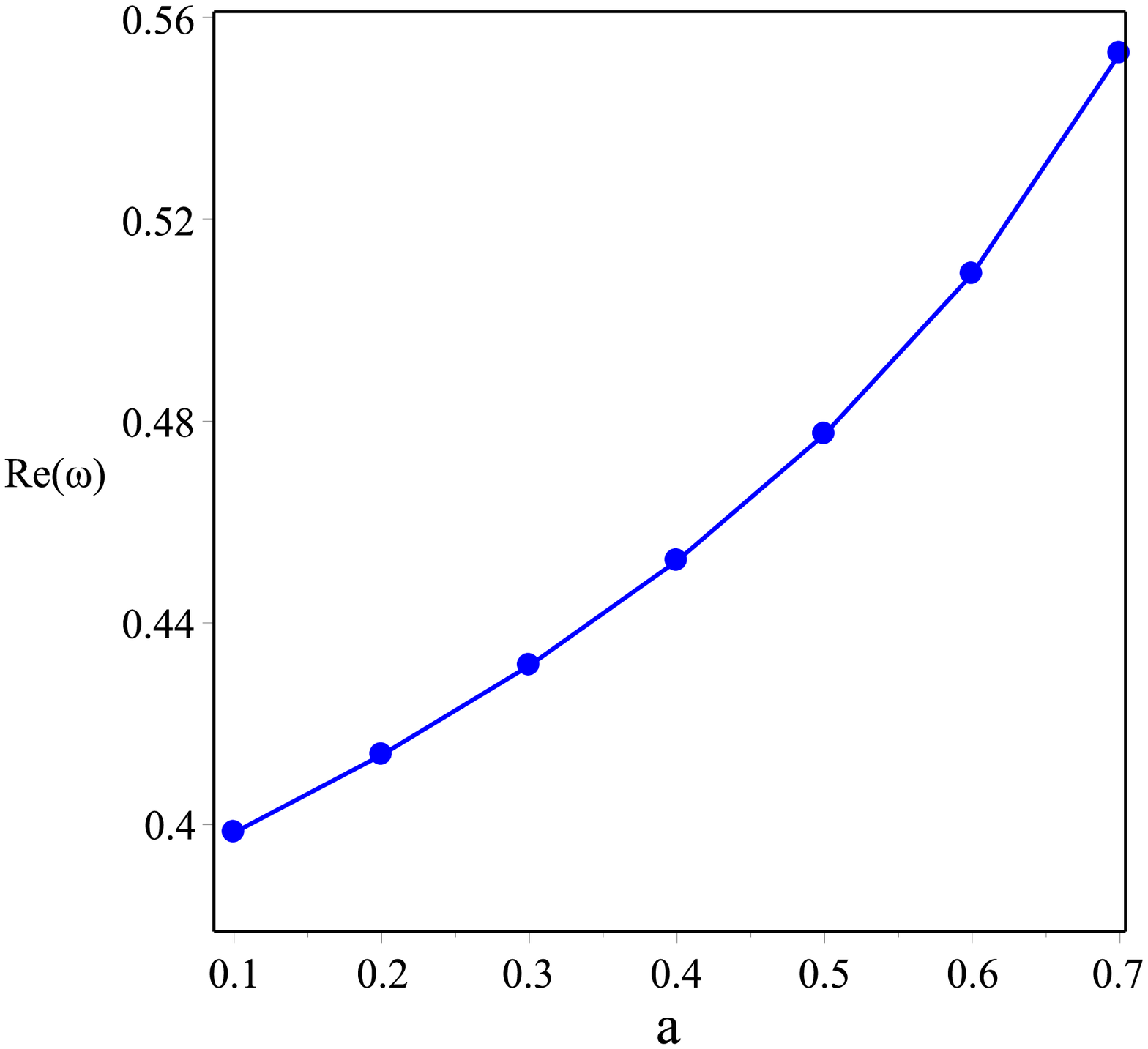} }\qquad \qquad \qquad \qquad 
    \subfigure[]{
   \label{FigQ2}  \includegraphics[width=0.33\textwidth]{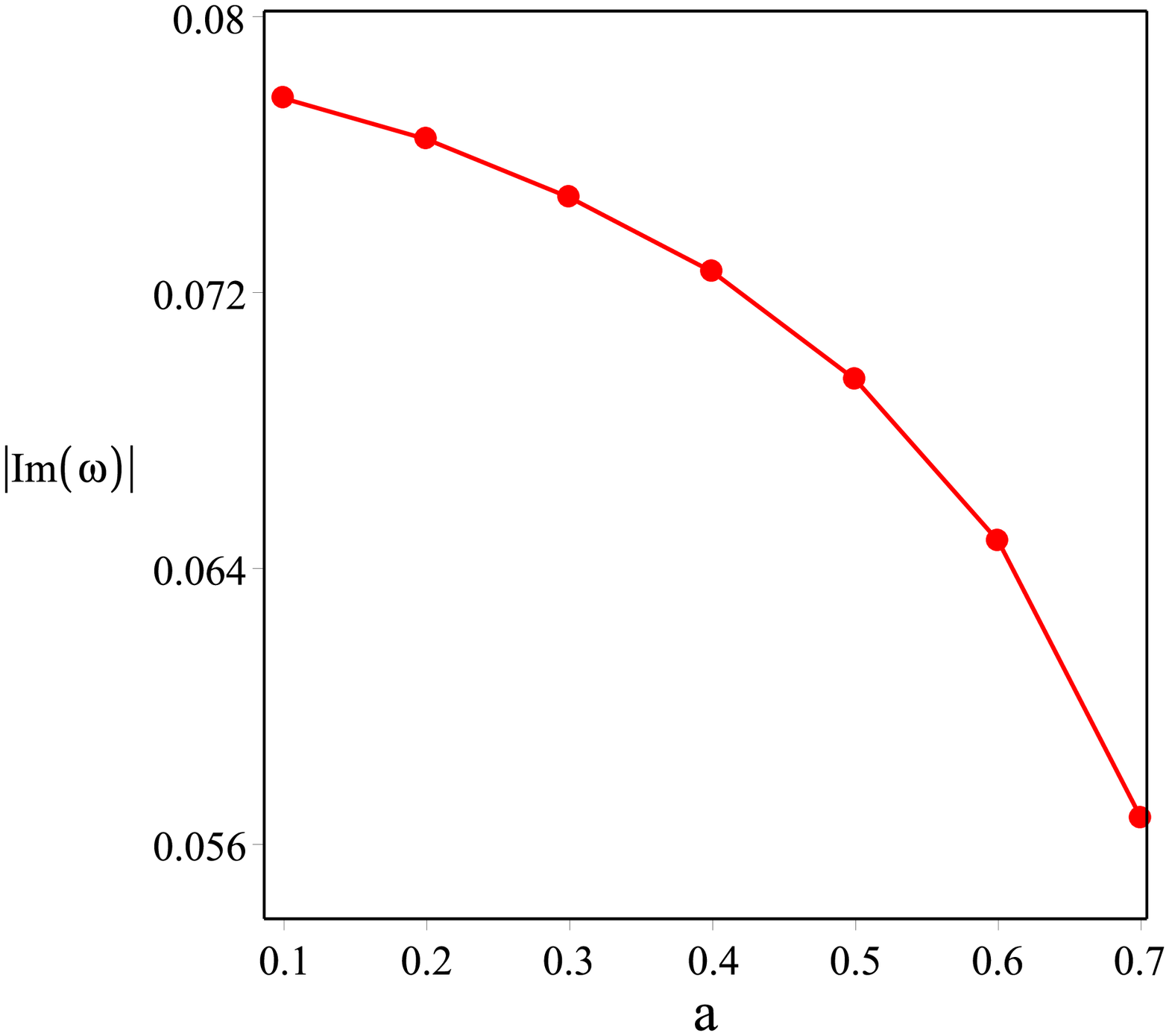}}
\caption{The behavior of Re($\omega$) and Im($\omega$) with respect to the suppression parameter for $ m=1$, $n = 0$ and $l = 2$.}
\label{Fig5}
\end{figure*}

\subsection{Quasinormal modes}

Now, we are interested in studying the QNMs around this kind of black holes with the help of Eqs. (\ref{EqQNM2}) and (\ref{EqQNM3}). The behavior of the real and the imaginary parts of the QNMs' frequencies versus the suppression parameter $a$ for RBHs with asymptotically Minkowski core is depicted in Fig. \ref{Fig5}. From Fig. \ref{FigQ1}, we see that the real part of the QN frequencies is an increasing function of $a$. This shows that the scalar field perturbations around the corresponding black hole oscillate with more energy for large values of this parameter. Taking a look at Fig. \ref{FigQ2}, one can find that the absolute value of the imaginary part of $  \omega$ reduces with the growth of $ a $. Since the inverse of $|\omega_{I}|$ determines the damping time, $t_{D} = |\omega_{I}|^{-1}$, one can say that  the scalar perturbations decay more slowly for the larger suppression parameter.

\section{Discussion and Conclusion}\label{V}

In this paper, we considered two interesting novel RBHs. One of them is a RBH characterised by a bounce parameter which shows an unusual feature of bouncing into a future incarnation of the universe rather than back into our own one. The other one is a structure with an exponentially suppressing mass (in the asymptotically Minkowski core region) controlled by a suppression parameter. 

First, we performed an in-depth study of the optical features on a charged black bounce and explored the impact of the bounce parameter and electric charge on them. The results indicated that although the radius of the photon sphere depends on both parameters $a$ and $Q$ for fixed mass, the shadow radius is independent of the bounce parameter and is just affected by the electric charge. Then, we continued by investigating the energy emission rate and examining the influence of the parameters on the radiation process. The results showed that as the effects of electric charge and bounce parameter get stronger, the the evaporation process gets slower. In other words, the lifetime of this black hole would be longer under such conditions.

Studying the mentioned optical properties of RBHs with the asymptotically Minkowski core, we found that both the photon sphere and the shadow radii depend on the model parameter, namely, the suppression parameter in this case and are decreasing functions of the parameter.
Regarding the effect of this parameter on the energy emission rate, our findings showed that the decreasing suppression parameter leads to a fast emission of particles. Thus, this black hole will have a short lifetime in the presence of a smaller suppression parameter.

Furthermore, we compared the resulting shadows to data of the EHT collaboration in order to find the allowed values of the model parameters. We found an upper limit for the electric charge so that the black bounce shadow has a size which is compatible with the EHT observations. Regarding the shadow size of the suppressed asymptotic Minkowski core regular black hole, we found that there is an upper limit for the suppression parameter for which the resulting shadow would be consistent with observational data.

Finally, we used the correspondence between the shadow radius and QNMs' frequencies and investigated scalar field perturbations  around the above-mentioned RBHs. Calculating the QNMs for RBHs with an asymptotically Minkowski core, we found that the scalar field perturbations around these black holes oscillate with more energy and decay slower for a larger suppression parameter. Regarding the charged black bounce, our results illustrated that the electric charge has an increasing effect on the real value of the QN frequency, whereas its effect on the imaginary part is dependent on the values of the bounce parameter. For large values of this parameter, Im($\omega$) is a decreasing function of $Q$, while for smaller values, Im($\omega$) grows up to a maximum value with increase of $Q$ and then gradually reduces as the electric charge increases more.

\begin{acknowledgments}
MKZ would like to thank Shahid Chamran University of Ahvaz, Iran for supporting this work.
FSNL acknowledges support from the Funda\c{c}\~{a}o para a Ci\^{e}ncia e a Tecnologia (FCT) Scientific Employment Stimulus contract with reference CEECINST/00032/2018, and funding from the research grants No. UID/FIS/04434/2020, No. PTDC/FIS-OUT/29048/2017 and No. CERN/FIS-PAR/0037/2019. 
\end{acknowledgments}


\end{document}